\documentstyle[11pt]{article}
\def\appendix{\renewcommand{\thesection}{\Alph{section}}\setcounter{section}{0}
              \renewcommand{\theequation}
            {\mbox{\Alph{section}.\arabic{equation}}}\setcounter{equation}{0}}
\def\maketitle{\thispagestyle{empty}\setcounter{page}0\newpage
                \renewcommand{\thefootnote}{\arabic{footnote}}
                  \setcounter{footnote}0}
\renewcommand{\thanks}[1]{\renewcommand{\thefootnote}{\fnsymbol{footnote}}
               \footnote{#1}\renewcommand{\thefootnote}{\arabic{footnote}}}
\newcommand{\preprint}[1]{\hfill{\sl preprint - #1}\par\bigskip\par\rm}
\renewcommand{\title}[1]{\begin{center}\Large\bf #1\end{center}\rm\par\bigskip}
\renewcommand{\author}[1]{\begin{center}\Large #1\end{center}}
\newcommand{\address}[1]{\begin{center}\large #1\end{center}}

\def\dinfn{\smallskip $^3$ Dipartimento di Fisica, Universit\`a di Trento\\ 
                           and Istituto Nazionale di Fisica Nucleare,\\
                                   Gruppo Collegato di Trento, Italia}

\def\Idinfn{\address{\dinfn}}
\def\dinbcn{\smallskip $^1$ Consejo Superior de Investigaciones
 Cient\'{\i}ficas \\
 IEEC, Edifici Nexus 201, Gran Capit\`a 2-4, 08034 Barcelona, Spain\\
 and Departament ECM and IFAE, Facultat de F\'{\i}sica, \\
 Universitat de Barcelona, Diagonal 647, 08028 Barcelona, Spain}
\def\Idinbcn{\address{\dinbcn}}
\newcommand{\email}[1]{e-mail: \sl #1@science.unitn.it\rm}
\newcommand{\zmail}[1]{e-mail: \sl #1@ieec.fcr.es\rm}
\newcommand{\fzmail}[1]{\thanks{\zmail{#1}}}
\newcommand{\femail}[1]{\thanks{\email{#1}}}
\newcommand{\pacs}[1]{\smallskip\noindent{\sl PACS numbers:
                       \hspace{0.3cm}#1}\par\bigskip\rm}
\def\babs{\hrule\par\begin{description}\item{Abstract: }\it}
\def\eabs{\par\end{description}\hrule\par\medskip\rm}
\renewcommand{\date}[1]{\par\bigskip\par\sl\hfill #1\par\medskip\par\rm}
\newcommand{\ack}[1]{\par\section*{Acknowledgements} #1}
\def\Idinbcn{\address{\dinbcn}}
\def\dimpcoll{\smallskip $^2$ Theoretical Physics Group, Imperial College, \\
Prince Consort Road, London SW7 2BZ, U.K.} 
\def\Idimpcoll{\address{\dimpcoll}}

\newcommand{\icmail}[1]{e-mail: \sl #1@ic.ac.uk\rm}
\newcommand{\ficmail}[1]{\thanks{\icmail{#1}}}
\def\eabs{\par\end{description}\hrule\par\medskip\rm}
\renewcommand{\date}[1]{\par\bigskip\par\sl\hfill #1\par\medskip\par\rm}
\renewcommand{\vec}[1]{{\bf #1}}       %%%  vectors in bold
\def\hs{\qquad}               %%%  horizontal space
\def\nn{\nonumber}            %%%  no number for eqnarray
\def\beq{\begin{eqnarray}}    %%%  begequation/eqnarray
\def\eeq{\end{eqnarray}}      %%%  endequation/eqnarray
\def\at{\left(}               %%%  open (
\def\aq{\left[}               %%%  open [
\def\ct{\right)}              %%%  close )
\def\cq{\right]}              %%%  close ]
\def\R{{\hbox{{\rm I}\kern-.2em\hbox{\rm R}}}}   %%% real numbers
\def\H{{\hbox{{\rm I}\kern-.2em\hbox{\rm H}}}}   %%% Hilbert space
\def\N{{\hbox{{\rm I}\kern-.2em\hbox{\rm N}}}}   %%% natural numbers
\def\C{{\ \hbox{{\rm I}\kern-.6em\hbox{\bf C}}}} %%% complex numbers
\def\Z{{\hbox{{\rm Z}\kern-.4em\hbox{\rm Z}}}}   %%% integers numbers
\newcommand{\fr}[2]{\mbox{$\frac{#1}{#2}$}}      %%% small fraction
\def\lap{\Delta}                                   %%% Laplacian
\def\be{\beta}

\def\de{\delta}

\def\ze{\zeta}

\def\la{\lambda}

\def\om{\omega}

\begin{document}
%\tableofcontents       %%%%%%   index of section

\preprint{Imperial/TP/97-98/36}

\title{Is the multiplicative anomaly relevant ?}

\author{Emilio Elizalde$^{1,}$\fzmail{elizalde}\,, \\
Antonio Filippi$^{2,}$\ficmail{a.filippi} \,, \\
Luciano Vanzo$^{3,}$\femail{vanzo} and
Sergio Zerbini$^{3,}$\femail{zerbini}
}
\Idinbcn
\Idimpcoll
\Idinfn

\date{April 1998}

\babs
In a recent work, S. Dowker has shed doubt on a recipe used in computing the 
partition function for a matrix valued operator. This recipe, advocated 
by Benson, Bernstein and Dodelson, leads naturally to the so called 
multiplicative anomaly for the zeta-function regularized functional 
determinants. In this letter we present arguments in favour of the
mentioned prescription, showing that it is the valid one in 
calculations involving the
relativistic charged bosonic ideal gas in the framework of functional 
analysis.

\eabs

\pacs{05.30.-d,05.30.Jp,11.10.Wx,11.15.Ex}

%\newpage

In a recent work, Stuart Dowker \cite{dowk98u} has shed doubt on a 
commonly used 
manipulation in computing functional determinants related to matrix 
valued elliptic operators. This recipe is widely used 
\cite{bens91-44-2480}--\cite{kirs97u-161} and can be summarized as follows.
Within the one-loop approximation or in the external field 
approximation, one often has to evaluate Euclidean functional integrals of the 
kind
\beq
Z=\int D\phi_1 D\phi_2 e^{-\int d^4x \phi_i A_{ij} \phi_j }
\:,\label{1}\eeq 
where $ A$ is a matrix valued differential operator, which we assume 
to have constant coefficients (this is not a restriction as long as 
one has to compute the one-loop effective potential). The result of the 
Gaussian functional integration is 
\beq
Z=\at \det A \ct^{-1/2}
\:.\label{2}\eeq
The problem is how to compute the functional determinant in 
Eq.~(\ref{2}).
Note that 
$A$ has matrix elements with discrete (field) and continuous (spacetime) indices.
If $A$ is also diagonal in the discrete indices, one has
\beq
 \hs A=\at \begin{array}{cc}
L_1 & 0 \\
 0  & L_2 
\end{array}
\ct
\:.\label{3}\eeq
 
In his work Dowker presents two different ways of computing this
determinant, claiming that one of them does not lead to correct
results.
The first recipe  \cite{bens91-44-2480} is equivalent to taking the
algebraic determinant first and the functional one afterwards, i.e. 
\beq
(\det A)_1=\det (L_1L_2)
\:.\label{44}\eeq
According to Dowker \cite{dowk98u}, it could seem more natural to use the 
alternative recipe 
\beq
(\det A)_2=\det L_1 \det L_2
\:.\label{dowk}\eeq
This is seen as the implementation of the right way to take the functional
determinant of the matrix valued  operator, i.e. an ordinary determinant
on both the spatial and field indices, at the same moment.
The example analized is that of two real free scalar fields of
different masses, for which the partition function has the form in 
(\ref{1}) and the operator is diagonal in the field indices $i,j$
as in (\ref{3}).
 
This second recipe seems quite natural indeed, since it is well known that
for finite matrices, $\det AB=\det A \det B$.  Unfortunately, in the
continuum, one needs a regularization and for regularized functional
determinants the two recipes may give different answers, because of
the existence of the so called multiplicative anomaly, discovered by
Wodzicki (see, for example, \cite{kass89-177-199}), 
\beq \det AB=\det A\det B e^{a(A,B)}
\:,\label{6}\eeq 
and, also in very simple cases of physical interest,
it is possible to show that $a(A,B)$ is not vanishing
\cite{eliz97u-394}. Thus the question posed by Dowker is substantial.

In our opinion, both recipes are formally acceptable. In fact
if one considers the eigenvalue problem for the matrix valued operator
$A$, one arrives at the formal determinant: $\det A =\prod_{n_1 n_2}
\la_{n_1} \la_{n_2}$, which can as well be rewritten as $\det A
=\prod_{n_1} \la_{n_1} \prod_{n_2} \la_{n_2}$. Of course, as
rigorous equalities these expressions are restricted to the finite
dimensional case (and to a limited class of absolutely convergent situations).

In the following we would like to present arguments in favour of the 
first recipe, analysing the general validity of both the recipes and
also  considering as a crucial example
the case of a free relativistic charged bosonic field 
at finite temperature \cite{kapu81-24-426,habe81-46-1497}. The 
self-interacting charged scalar field was studied in 
\cite{bens91-44-2480}.

We have argued that both recipes actually coincide
in the finite dimensional case. One should keep in mind that any
extension to the continuous,
functional case always starts from a discretization. In particular, the
finite dimensional example posed by Dowker at the end of \cite{dowk98u}
pretending to prove a discrepancy already at this level does not 
apply.
In fact, let us start considering Dowker statements in general.
He supports the inequivalence of the two recipes
analysing a generic four by four matrix $A_{\alpha\beta i j}$
diagonal on $i,j$, where the indices $\alpha , \beta =
1,2 $ represent the discretized version of the continuous space-time
indices and $i,j=1,2$ the fields indices.

Then, the two recipes correspond to $\det_{\alpha,\beta}\det_{i,j}
A_{\alpha\beta i j}$ and  $\det_{\alpha,\beta,i,j}
A_{\alpha\beta i j}$ respectively.
For such a generic matrix it is straightforward to see that the two results
are different and Dowker's statement  that the first recipe could
give inexact results seems correct.
The point, though, is that the matrices we encounter in field theory have
additional structural requirements. In the continuous limit, the matrix
valued differential operator is normally diagonal in the continuous
indices, namely
\beq
A_{i,j}(x,y)\equiv A_{i,j} \delta(x,y) 
\:.\label{7}
\eeq
and will therefore have a  block-diagonal structure.
The determinant of such a matrix is the product of
the determinants of the blocks, therefore
\beq
\det_{\alpha,\beta,i,j}
A_{\alpha\beta i j}= \det_{\alpha,\beta}\det_{i,j} A_{\alpha \beta i j},\label{equal}
\eeq
which is exactly the widely used recipe.

A more formal analysis would require the study of the proper discrete
matrix, for the functional integral is solely defined as the continuum limit
of a discretized lattice version. The derivative there is  in fact defined
as difference of the values of the field in two neighbouring points of the
lattice and the derivative squared has, therefore,
 terms which are off-diagonal in
the spacetime indices, like $\phi(x+1)\phi(x)$.
In one dimension the corresponding matrix $A_{\alpha \beta i j}$ would have the structure
\beq
 \hs A=\at \begin{array}{ccccc}
 \Box &\Diamond  &0 & 0 & \cdots\\
 0  &\Box &\Diamond  & 0 & \cdots\\
 0 & 0 & \Box & \Diamond & \cdots\\
 0 &0 & 0 & \Box & \cdots\\
 \vdots &\vdots & \vdots & \vdots   & \ddots  \\
\end{array}
\ct
\:.\label{8}\eeq
where $\Box$ and $\Diamond$ represent blocks in the field indices. 
It is easy to see how, for a  matrix with such a structure, the above
equality (\ref{equal}) holds again, since the only contributions to
the determinant come from the diagonal blocks. 

It thus seems reasonable that the first, widely used
recipe for calculating the functional determinant is equivalent to the
other more rigorous one in the {\it finite limit} of field
theory. This shows also that the reasons for the presence of the anomaly have
to be found in the infinite nature of the functional determinant.

Let us turn our attention to the free charged bosonic field at finite 
temperature.  
In order to evaluate functional determinants we will make use of 
zeta-function regularization
\cite{dowk76-13-3224,hawk77-55-133}. 
Recall that the one-loop Euclidean partition 
function, regularized by means of 
zeta-function techniques, reads \cite{hawk77-55-133}
\beq
\ln Z=-\frac{1}{2}\ln\det \at L_D \ell^2 \ct
=\frac{1}{2}\ze'(0|L_D)-\frac{1}{2}\ze(0|L_D)\ln \ell^2
\nn\:,\eeq
where $\ze(s|L_D)$ is the zeta function corresponding to $L_D$,
a second order elliptic differential operator,
$\ze'(0|L_D)$ its derivative with respect to $s$, and
$\ell^2$ is a renormalization scale parameter.

The grand canonical partition function for an ideal charged
relativistic boson gas  may be written as ($\mu$ is the chemical
potential) \cite{bens91-44-2480}:
\beq
Z_{\be,\mu}=\int_{\phi(\tau)=\phi(\tau+\be)}D\phi_i
e^{-\fr{1}{2}\int_0^\be d\tau \int d^3x d^3y
\phi_i(x)A_{ij}(x,y)\phi_j(y)}
\:,\label{}\eeq
where the two real fields $\phi_i$, $i=1,2$,  are chosen to describe the
degrees
of freedom of the boson gas, and the operator $A$ has  matrix elements
given by
\beq
A_{ij}(x,y)= \at L_{ij}+ 2\mu \epsilon_{ij}
\sqrt{L_\tau} \ct \delta(x,y)
,\label{sd}\eeq
with
\beq
L_{ij}=\at L_\tau+ L_3-\mu^2
\ct\de_{ij},\,\,\, L_3=-\lap_3+m^2
\:,\label{nm}\eeq
in which $\lap_3$ is the Laplace operator on $\R^3$ (continuous
spectrum $\vec k^2$) and
$L_\tau=-\partial^2_\tau$ (discrete spectrum
$\om_n^2=\frac{4\pi^2 n^2}{\be^2}$) the Laplace operator on $S^1$. In
this case,  the partition function may be written as \cite{bens91-44-2480}
\beq
Z_{\be,\mu}=\at 
\det \left\{ \ell^2 \at \begin{array}{cc}
L_\tau+ L_3-\mu^2 & 2\mu \sqrt{L_\tau}\\
-2\mu \sqrt{L_\tau} & L_\tau+ L_3-\mu^2
\end{array}
 \ct  \right\}\ct^{-1/2}
\label{o2}
\eeq
The  first recipe consists in taking first the algebraic determinant  and
then the functional determinant. The result is
\beq
Z_{\be,\mu}= \at \det \left\{ \ell^4 \aq (L_\tau+ L_3-\mu^2)^2+4 \mu^2
L_\tau \cq \right\}\ct^{-1/2} = \at \det ( \ell^4 L_+L_-)\ct^{-1/2}
\:,\label{05}\eeq
where
\beq
L_\pm=L_\tau + L_3+\mu^2 \pm 2 \mu \at L_3 \ct^{\fr{1}{2}}
\:.\label{bbd}
\eeq
In an attempt of using the second recipe, one may observe that  every  $2 \times 2$ matrix
here can always be diagonalized. Then, modulo a trivial
functional Jacobian corresponding to the diagonalization,  one has
\beq
Z_{\be,\mu}=\at 
\det \left\{ \ell^2 \at \begin{array}{cc}
L_+ & 0\\
0 & L_-
\end{array}
 \ct  \right\}\ct^{-1/2}
\label{o289}
\eeq
and the answer coming from the second recipe is
\beq
Z_{\be,\mu}=\at \det ( \ell^2 L_+) \det ( \ell^2 L_-)\ct^{-1/2}
\:.\label{051}\eeq

On the other hand ---as is usually done in quantum field theory--- one
can also
describe the gas by two complex fields $\phi$ and $\phi^*$, defined by
\beq
\phi=\frac{1}{\sqrt 2}\at \phi_1+i \phi_2 \ct\,\,\hs
\phi^*=\frac{1}{\sqrt 2}\at \phi_1-i\phi_2 \ct
\:.\label{bo11}\eeq
The corresponding grand partition function reads \cite{kapu81-24-426}
\beq
Z_{\be,\mu}=\at
\det \left\{ \ell^2 \at \begin{array}{cc}
K_+ & 0\\
0 & K_-
\end{array}
 \ct \right\}\ct^{-1/2}
\label{o22}
\eeq
where now
\beq
K_\pm=L_3+L_\tau-\mu^2 \pm 2i\mu \at L_\tau
\ct^{\fr{1}{2}}
\:.\label{z}\eeq

Note that we have  $L_+L_-=K_+K_-$, thus the first recipe gives the same
answer for the two approaches, namely
\beq
Z_{\be,\mu}= \at \det \aq \ell^4 (L_+L_-) \cq \ct^{-1/2}=\at \det \aq \ell^4 (K_+K_-) \cq
\ct^{-1/2}\:.\label{r1}\eeq
The second recipe gives, on its turn,
\beq
Z_{\be,\mu}= \at \det \at \ell^2 L_+ \ct \det \at \ell^2 L_- \ct \ct^{-1/2}= 
\at \det
\at \ell^2 K_+\ct  \det \at \ell^2 K_-\ct \ct^{-1/2} \,.
\label{nmv}
\eeq
When the multiplicative anomaly is non vanishing, one of two recipes
is in contradiction. In Ref. \cite{eliz97u-404} it has been shown
that in odd dimensions, the multiplicative anomaly is vanishing
and the two recipes give the same answer. In even dimensions, in
particular in $\R^4$, the multiplicative anomaly is non vanishing and
only the first recipe gives the same partition function for the two 
approaches, since we
have
\beq
\ln \at \ell^2 \det L_+\ct +\ln \at \ell^2 \det L_-\ct +a(L_+,L_-)=\ln
\at \ell^2 \det K_+\ct +\ln \at \ell^2 \det K_-\ct +a(K_+,K_-)
\label{bbb}\eeq
but, on the contrary,
\beq
\ln \at \ell^2\det L_+\ct +  \ln \at \ell^2 \det L_-\ct  \neq \ln \at
\ell^2 \det K_+\ct +\ln \at \ell^2 \det K_-\ct
\label{bbbbb}
\,.\eeq
It  has also  been shown that the statistical sum contribution to the 
grand partition function is the 
same and yields the well known expression \cite{habe81-46-1497}
\beq
S(\be,\mu)=\sum_i \aq \ln \at 1-e^{-\be(\sqrt \la_i+\mu)} \ct+\ln \at 
1-e^{-\be(\sqrt \la_i-\mu)} \ct \cq
\:,\label{ss}\eeq
where $\la_i\equiv \vec k^2+m^2$. The discrepancy manifests itself in the 
"vacuum sector", namely in the contribution to the grand partition 
function linear in $\be$ and the presence of the multiplicative 
anomaly, first recipe, renders the grand partition functions 
equal and independent by the parametrization of the degrees of freedom 
of the relativistic ideal gas.  

Having performed the above calculation ---that seems to leave
little chance for discrepancy--- one might still ask: how could the
apparently more
natural second recipe Eq. (\ref{dowk}) fail~?  Well,
the answer is, to begin with,
that we are dealing with a very elusive mathematical and physical
point. In fact, for a {\it direct sum} of operators, acting on a
corresponding direct sum of independent functional spaces 
(this is the case in quantum physics
when, for instance, a superselection rule is imposed upon the system),
we indeed have the factorization property:
\beq
 \det A=\at \begin{array}{ccc}
A_1 & 0 & \cdots \\
 0  & A_2 & \cdots \\
 \vdots  & \vdots & \ddots
\end{array}
\ct  = \det A_1 \det A_2 \cdots
\:.\label{e1}\eeq
In this situation the product itself of the operators, $A_1A_2$, makes
no sense in general, much less its determinant,
 and the prescription $\det A_1 \det A_2 \cdots$ is to be used.
 But things
are absolutely different when one is working within a functional space
where field mixing and rotations are allowed,
 and one just obtains the diagonal expression for the
matrix of operators acting in this space as a particular form, 
after a convenient diagonalization process
(this is precisely what happens in our example above). 
The moral we extract from the outcome of our analysis is that 
the {\it invariant} which is preserved under
this process of change of basis is precisely {\it the determinant of the
operator matrix},
but not the fact that it is equal to the product of the functional determinants
of the operators. It is precisely that
invariance what lays in the heart of our example: the noncommutative
anomaly is the missing term necessary in order to preserve it.

To summarize, we have here carried out what is, in our opinion, a serious
consistency check in favour of the first
recipe for the calculation of determinants of  matrix valued operators  and,
as a consequence, provided arguments in favour of the relevance of  
zeta-function regularization and related multiplicative anomaly in quantum field theory. 
As far as this last issue is concerned, in Ref. 
\cite{eliz98u-1} we also  respond to  a criticism appeared recently 
\cite{tse98}, concerning the use of zeta-function regularization.

\ack{We would like to thank G. Cognola, V. Moretti, R. Rivers and T. Evans 
 for valuable discussions. 
This work has been supported by the cooperative agreement
INFN (Italy)--DGICYT (Spain).
EE has been partly financed by DGICYT (Spain), project PB96-0925, and
by  CIRIT (Generalitat de Catalunya),  grant 1995SGR-00602. AF wishes 
to acknowledge financial support from the European Commission under TMR 
contract N. ERBFMBICT972020.
}


\begin{thebibliography}{10}


\bibitem{dowk98u}
J.S.~Dowker. 
{\em On the relevance of the multiplicative anomaly},
Preprint MUTP/98/4, hep--th/9803200


\bibitem{bens91-44-2480}
K.~Benson, J.~Bernstein and S.~Dodelson.
 Phys.~Rev. {\bf D 44}, 2480 (1991).


\bibitem{kapu81-24-426}
J.I.~Kapusta.
 Phys.~Rev. {\bf D 24}, 426 (1981).

\bibitem{kapu93b}
J.~Kapusta.
{\em Finite-temperature field theory},
Cambridge University Press, Cambridge,(1993).

\bibitem{toms93}
D.J.~Toms. Phys. Rev. Letts.  {\bf 69},  
1152,  (1992); Phys. Rev.  {\bf D 47},  
2483,  (1993).


\bibitem{kirs95-51-6886}
K.~Kirsten and D.J.~Toms. Phys. Rev.  {\bf D 51},  
6886,  (1995).

\bibitem{kirs97u-161}
K.~Kirsten and D.J.~Toms. Phys. Lett. {\bf B 368}, 119, (1996);
Phys. Rev.  {\bf D 55}, 7797, (1997).


\bibitem{kass89-177-199}
C.~Kassel.
Asterisque {\bf 177}, 199 (1989), Sem.~Bourbaki.


\bibitem{eliz97u-394}
E.~Elizalde, L.~Vanzo and S.~Zerbini,  {\em Zeta-function Regularization, the
  Multiplicative Anomaly and the Wodzicki Residue},  hep--th/9701060
(1997), to appear in Commun.~Math.~Phys.


\bibitem{habe81-46-1497}
H.E.~Haber and H.A.~Weldon.
Phys.~Rev.~Lett. {\bf 46}, 1497 (1981).


\bibitem{dowk76-13-3224}
J.S.~Dowker and R.~Critchley.
 Phys.~Rev. {\bf D 13}, 3224 (1976).

\bibitem{hawk77-55-133}
S.W.~Hawking.
 Commun.~Math.~Phys. {\bf 55}, 133 (1977).

\bibitem{eliz97u-404}
E.~Elizalde, A.~Filippi, L.~Vanzo and S.~Zerbini,  {\em One-loop 
Effective Potential for a Fixed Charged Self-interacting Bosonic 
Model at Finite Temperature with its Related  
Multiplicative Anomaly },  hep--th/9710171 (1997), to appear in
Phys.~Rev.~D.

\bibitem{eliz98u-1}
E.~Elizalde, A.~Filippi, L.~Vanzo and S.~Zerbini,  {\em  
Is the multiplicative anomaly dependent on the regularization~? },  
hep--th/9804071 
(1998).

\bibitem{tse98}
T.S. Evans.
{\em Regularization schemes and the multiplicative anomaly},
 hep--th/9803184.


\end{thebibliography}
\end{document}